\documentclass[twocolumn]{revtex4-1}
\usepackage{amssymb}
\usepackage{graphicx}
\usepackage{latexsym}
\begin{document}
\title{ Adiabatic index for relativistic stars as a coefficient of 
non-thermal dissipation }
\author{ Seema Satin }
\affiliation{ Indian Institute for Science Education and Research , Pune 
 , India.}
\email{seemasatin@iiserpune.ac.in}
\newcommand{\be}{\begin{equation}}
\newcommand{\ee}{\end{equation}}
\newcommand{\bea}{\begin{eqnarray}}
\newcommand{\eea}{\end{eqnarray}}
\begin{abstract}
The adiabatic index of a relativistic star (modeled by a perfect fluid)
 is shown to act as a dissipation constant due to considerations of 
mesoscopic scale stochastic effects. This dissipative effect   
  arises without taking heat flux in the fluid model and is 
termed as adiabatic or non-thermal for the system under consideration. 
A basic formalism for introducing stochastic effects in interiors of massive 
stars has recently been proposed via a classical Einstein-Langevin equation. 
 The origin of stochasticity is associated with the pressure variable (due to
 degeneracy of constituent particles ) and its fluctuations,
 and is not viable for pressureless fluids.
 The fluctuations dissipate their fermi energy into perturbing the system,
 adiabatically. This is shown by a fluctuation dissipation relation, and
  a Langevin fermi temperature associated with these fluctuations is defined.
The high magnitude of background field fluctuations which we get here, is
 feasible to account for perturbing the strong gravity regions with dense
 matter. Nevertheless, the perturbations themselves are consistently and
 reasonably small as first order deviations from equilibrium.     
\end{abstract}

\maketitle
Exploring the interiors of relativistic/compact stars and the properties of
 matter in highly dense state of matter in strong gravity regions, poses a
 challenge for astronomers as well as for theoretical physicists for modeling
 them \cite{norman, lattimer, pines}. 
 With the advent of gravitational wave detection methods 
\cite{pratten, maggie}, one can  probe into these, by developing more detailed
 formalism that would help  to delve into deeper layers in one way or the
 other . 
Hence this area is open for enhancing the basic models and extending towards  
  new methods of analysis in order to decipher the interiors of such 
astrophysical objects, theoretically as well as for observational purposes.
  In this letter we show that the adiabatic index can be
identified as a coefficient of non-thermal dissipation for cold
matter relativistic star models, enhanced with stochastic effects. These 
stochastic fluctuations which partially capture the microscopic/quantum 
properties resulting in a mesoscopic description of the  matter in the 
interiors, act as source of induced perturbations. It is not
  just the magnitude of these mesoscopic fluctuations
that is of interest  to emphasise the importance of such additions, but
new underlying physics which can be explored due to such considerations 
is the aim of such an endeavour.
 The basic structure of such a theory begins with  a classical
 Einstein-Langevin formalism in recent work \cite{seema1, seema2}. 
  While in this letter we obtain a fluctuation-dissipation
relation and give  estimates over magnitude of background fluctuations for a
 simple model, which are obtained due to known approximate values of the 
adiabatic index of the neutron star.
This also opens up new avenues in statistical physics, both equilibrium and
non-equilibrium, to extend such developments and studies for compact stars and
objects governed by general relativity.

 Considering a zero temperature
 equation of state to describe matter $\epsilon = \epsilon(p) $, say for a
 neutron star ( one-component perfect fluid ) where pressures are quantum 
mechanical in origin due to degeneracy of neutrons in order to encounter strong
 gravity, we model noise in the system. The stress tensor for such a system
is that of a the perfect fluid,
\be
 T^{ab}  =  u^a u^b \epsilon + q^{ab} p 
\ee
where $ q^{ab} = u^a u^b + g^{ab} $.
  
The identification of the adiabatic index of the massive star, given by
\be \label{eq:ad}
\gamma = \frac{ (\epsilon + p) }{p} \frac{dp}{d \epsilon }   
\ee
and  defined as the fractional change in pressure per fractional change in 
comoving volume at constant entropy and composition for a massive star, with 
 a dissipative term is seen to follow from the F-D  relation that we obtain
 here.
It is known that coefficient of friction/dissipation in thermal systems,
 depends on details of molecular size and their kinetic energy in the medium 
and is linked to temperature and pressure of the system \cite{kubo} . However
 it is not
a thermodynamic property, and can  be linked to other parameters than 
the thermal temperature of the system.  Here we connect  the 
adiabatic index as a coefficient of non-thermal dissipation with the 
background noise (fluctuations)  due to degeneracy pressure of neutrons.
 The fermi energy which is associated with the degeneracy pressure in  
The neutron stars  after they cool down to have thermal temperatures 
$\sim$ 1 MeV, while the fermi temperature is more than 
 $ E_F \approx $ 60 MeV \cite{fried}. 
The  stochastic effects which we model, thus encode these effects of quantum
 origin and are not thermodynamical. 

However since the equation of state is not known correctly for such cases, we
 do not claim any new estimates over magnitudes of fermi temperature or
 polytropic index, but work with known estimates only, introducing in the 
model these additional effects in terms of fluctuations.  
 
On the other hand, study of oscillations and perturbations in relativistic 
stars, forms the base
 of asteroseismology and theoretical developments start with perturbed
Einstein's equations \cite{shutz,fried}.
\be
 \delta G^{ab}(x) = 8 \pi \delta T^{ab}(x)
\ee
 However all through the
 well established literature, the perturbed Einstein's equations lack a source
 term, which may induce such perturbations in the system. The perturbations in
 an isolated  relativistic star or a more involved  configuration of 
binary/accreting systems are usually assumed to arise due to internal or 
external effects \cite{ander} which are often overlooked and not specified.
 The focus is
 rather over calculating and classifying the perturbations and modes of 
oscillations in an astrophysical system \cite{kokkotas}. Taking into 
consideration a source which induces perturbations in the system, is
 desirable, specially when these 
sources are due to some mechanism in the interior of the star itself.
These could also arise as response to some external agency which triggers 
internal mechanism in the matter content to start off the oscillations. 
 The characteristics of the perturbations depend on the properties of the 
stellar structure, rather than the details of the external source.  
 
 Our analysis is useful for cold neutron stars and other compact matter
 objects in a different way than just calculating estimates over TOV equations
 and mass radius relation, though exploring these could also be of significance
 in some specific cases.
 It is thus useful to lay a new strategy to explore the interiors of these
 objects based on enhancement of fluid properties as we do here.

 In this letter we state the basic framework for radial perturbations
  which has been worked out in \cite{seema1,seema2}, and  show 
the first  example of a parameter namely the adiabatic index that can be
viewed in a new way.

 The perturbations are regarded as deformation of the
 original configuration of the star, with each fluid element 
mapped to a corresponding fluid element by a Lagrangian displacement vector
$\xi^a$ and by metric perturbation $h_{ab}$. Thus fundamental variables
 describing the perturbations are $\xi^a$ and $h_{ab} $ \cite{fried}.
 With a source term acting as driving mechanism ( in the background spacetime)
 to induce  stochastic perturbations, the classical Einstein-Langevin equation
 reads \cite{seema1},
\be \label{eq:pee}
\delta G^{ab}[h,\xi;x) = 8 \pi (  \delta T^{ab}[h,\xi;x)  + \tau^{ab}[g;x) )
\ee
where the Langevin noise $\tau^{ab}[g;x) = T^{ab}[g;x) - < T^{ab}[g;x)> $ is
 defined on the background metric $g_{ab} $, and satisfies $<\tau^{ab}(x)> =0$.
The two point noise  kernel is given by $ <\tau^{ab}(x) \tau^{cd}(x')> =
 N^{abcd}(x,x')$, a bitensor. Such a treament of the gravitating system, calls 
for a  system-bath  separation. One can consider the spacetime metric
to act as the system while the matter fields as the environment/bath.

 Modes of oscillations which can be detectable 
from such isolated objects for realistic cases, and originating in the 
interiors can travel through the compact object to the surface and
further , in principle, in the form of gravitational waves, for non-radial
perturbations.  
 It is possible that these stochastic effects  set stage for new stochastic
modes of oscillations and finding them. 
 However one has to correctly work out the  corresponding analysis and show 
details of such  modes.  This can be carried 
out in detailed upcoming work.

The  pressures inside neutron stars are of the order of $10^{31}  - 10^{34}$ $
 Pa $ and they increase from the outer radius to the core modeled by perfect
fluids. 
 A correspondence between quantum field fluctuations in classicalized  limit
to hydrodymanic approximation in terms of macroscopic variables of the perfect
 fluid,  has recently been obtained in \cite{seema3}. Thus one can
 see, how any quantum fluctuations can be partially captured via a fluid 
model approximation of matter.    

For a relativistic star, the equation of state in simplest form is
\be
 p = w \epsilon^\Gamma
\ee
where $\epsilon$ is the energy density of the fluid, and $\Gamma$ the
 polytropic index. We give the above relation just for the sake of completion
of describing the system, and the results are not affected by unsettled issues
over polytropic index  for neutron stars in this letter.  One may as well
consider a simple case of $p = w \epsilon$ here for a toy model that can be 
used for such a relativistic star, as is used in literature for 
complete gravitational collapse of massive stars to singularities 
\cite{misner}.
It is apparent that the pressure fluctuations extend the stochastic nature 
to the energy density  via the equation of state. However we assign the origin
 of fluctuations in the pressure variable, as it is pressure in the neutron
 star which balances gravity.
 One  can consistently add the fluctuations of the stress
 tensor given by $\tau^{ab}$ as Langevin noise in the system. It is to be noted
that the background spacetime  with $g_{ab}$ is treated
 deterministically  while the perturbations  $ h_{ab} , \xi^a $ show a 
stochastic nature, hence $G^{ab}[g] =  8 \pi T^{ab} [g]$  still remains the
unaltered Einstein's equation while its perturbations 
$ < \delta G^{ab}>   = 8 \pi <\delta T^{ab}> $ are stochastic and have an
 added noise  on the rhs to give it a Langevin form. The perturbed metric 
is then of the form  $ g_{ab} + <h_{ab}>_s , < >_s $ being the stochastic
average.

For the present analysis we restrict ourselves to linearized equations, thus
limiting the perturbations of the Einstein's equations to first order in 
$h_{ab}$ and $\xi^a$. This can be obtained by second order variation of the
 action 
\be
 S = - \int (\frac{1}{16 \pi }R - \epsilon) \sqrt{g} d^4x
\ee
and is well established in literature, giving the perturbed Einstien's
 equation (\ref{eq:pee}), to which
we add the Langevin term and write the  E-L  equation  with explicit form
of $\delta T^{ab} $ as  
\be \label{diss}
 \delta G^{ab} = 8 \pi ( W^{abcd} \Delta g_{cd}
 - \mathcal{L}_\xi T^{ab} ) + 8 \pi \tau^{ab}
\ee
where $\Delta g_{ab} = h_{ab} + \nabla_a \xi_b + \nabla_b \xi_a $. 
The Lagrangian change in a quantity is given by $\Delta $ , while Eulerian
changes are respresented by $\delta$, the usual convention followed by
\cite{shutz,fried} 
We split and write the term 
\bea
& & W^{abcd}  =  E^{abcd} + D^{abcd} \\
 \mbox{ with }    \nonumber \\
& & E^{abcd}  = \frac{1}{2} ( \epsilon + p) u^a u^b u^c u^d +  \nonumber \\
& & \frac{1}{2} p ( g^{ab} g^{cd} - g^{ac} g^{bd} - g^{ad} g^{bd} ) 
 \nonumber  \\
& &  D^{abcd}  =   - \frac{1}{2} \gamma p q^{ab} q^{cd}  
\eea
here $\gamma$ stands for the adiabatic index  of the star as defined in
equation (\ref{eq:ad}). 
  We claim that the term $D^{abcd}$ shows a dissipative effect
and in absence of any heat flux. 
The claim is based on the reasoning that, since  noise here is sourced in
pressure fluctuations of the fluid fundamentally, the dissipative effect 
would be seen in the pressure perturbations $\Delta p = - 1/2 \gamma p q^{cd}
\Delta g_{cd} $ which in the perturbed stress tensor accounts for the 
$q^{ab} \Delta p $ term only, contributing to  $D^{abcd} \Delta g_{cd} $
 in the E-L equation. 
  More specifically,  the degeneracy
pressure fluctuations in the compact object dissipate their fermi (non-thermal)
energy into perturbing the background fluid variables and the geometry
given in terms of $ \Delta g_{ab}$.  
 Also $\Delta g_{cd} $  carries a harmonic time dependence  here 
and hence we get a local dissipation effect, for the temporal coordinate.

 The nature of noise in the static background model is such 
that stochasticity is seen prominently in the radial coordinate,  which means 
pressure fluctuations vary randomly with some distribution along the
radial vector, for the spherically symmetric geometry containing the
 perfect fluid.
\be \label{eq:le}
 ds^2 = - e^{2 \nu(r)} dt^2 + e^{2 \lambda(r)} + r^2 (d \Omega^2)
\ee

  Thus the radial dependence of noise is an important 
point to note here in the Langevin term, for different layers in the star.
It is this nature of noise (having radial dependence) at different depths,
 that such models can  help to probe the interiors in a 
compact star.  The temporal dependence of the terms is taken harmonic  (as in
\cite{seema1}) for this static background configuration for simplicity.  

  We  associate the above said dissipative term with the noise in the system
 having an F-D relation as discussed further.
   
The general form of a fluctuation dissipation relation, is known to be given as 
\be \label{eq:f}
 N^{abcd}(x,x') = \int K(x,x'') \tilde{D}^{abcd}(x'',x') d^4 x''
\ee 
where $N^{abcd}(x,x') $  and $ \tilde{D}^{abcd}(x'',x') $ are the noise
and dissipation kernels respectively,  $K(x,x'') $ being the
fluctuation-dissipation kernel. 
 Such a general form of FD theorem also follows for semiclassical stochastic
Einstein-Langevin equations \cite{bei1,bei2} though the details are obtained
 through formulations specific to quantum open systems.
 However the above relation is valid for classical
 systems as well, relating the fluctuations in a system to the dissipation.

For the astrophysical model that we consider here with  
metric given by  equation (\ref{eq:le}) and a perfect fluid stress tensor,
 the noise kernel  has been defined and obtained in all its details in
 \cite{seema1}, which is of the form  
\be
N^{abcd}(x,x') = \mbox{Cov}[T^{ab}(\vec{x}),T^{cd}(\vec{x}')] e^{-i \omega 
(t-t')}
\ee
with fluctuations  $\tau^{ab}(x) \equiv (T^{ab}(\vec{x}) -
 <T^{ab}(\vec{x})> ) e^{i \omega t } $.
The fundamental elements of perturbations are given by
 $ h_{ab}(x) \equiv h_{ab}(\vec{x}) e^{i \omega t} $ and $ \xi^a(x) \equiv
\xi^a (\vec{x}) e^{i \omega t} $, oscillating with 
frequency $\omega $ over the static background. 
The components of the point separated (two point) noise kernel, as one can 
 see in the  above reference, are  composed of two point covariances of 
pressure and energy density. In the coincidence limit
of spatial points , where we take $\vec{x}' \rightarrow \vec{x} $
 these reduce to the variances of pressure and energy density, keeping the
 temporal dependence separate. 

As one can clearly  see from the above relations, dissipation kernel 
 $D^{abcd}(x) = D^{abcd}(\vec{x})$, is local in spatial coordinates,
 for the static background fluid,
 while the factor $\Delta g_{cd} $ carries the perturbations and also the time
 dependence. This 
leads to ascertain taking coincidence limit in spatial coordinates 
for the two point noise kernel, with $\vec{x} \rightarrow \vec{x}' $. Denoting 
$(t-t') \equiv \mathbf{T}$, the F-D theorem here takes the form
\be
N^{abcd}( \vec{x})e^{-i \omega \mathbf{T}} = \int K(\vec{x } , \vec{x}'; 
\mathbf{T}) D^{abcd}(\vec{x}') d^3 \vec{x}'  
\ee
The simplest  choice for the fluctuation- dissipation kernel is
  $ K( \vec{x}, \vec{x}', \mathbf{T}) =  -\frac{<p(x)>}{2}
 \delta ( \vec{x}, \vec{x}') e^{-i \omega \mathbf{T}}$.

For the given case $q^{ab} = 0$ when $\{ab \}  = \{tt,tr,t\theta,t\phi\}$
(since $u^a \equiv \{ e^{-\nu},0,0,0 \} $) and $q^{\alpha \beta } = 
g^{\alpha \beta}$ for indices spanning over $ \{r,\theta,\phi\} $.  
It then follows that $D^{tttt} = D^{tt\zeta \delta}= 
D^{\alpha \beta tt} =  0 $ (where $\alpha, \beta,\zeta,\delta $ span over 
 spatial coordinates only), which means there is no dissipation in the
system with the corresponding noise components.
 The non-zero components of dissipation are then  given by 
$D^{\alpha \beta \zeta \delta}(\vec{x}) $, and one can write the F-D relation 
as 
\be \label{eq:nd}
 N^{\alpha \beta \zeta \delta }(\vec{x}) e^{-i \omega \mathbf{T}}  = - 
 \frac{<p(x)>}{2} D^{\alpha \beta \zeta \delta}(\vec{x})  e^{-i\omega 
\mathbf{T}} \nonumber 
\ee
which reduces to 
\be
 \mbox{Var[p(r)]}  =    \frac{1}{4} \gamma <p(r)>^2 
\ee
  The factor $\gamma/4$ here, acts as coefficient of adiabatic dissipation and
 has values $0.33$ and $0.41$ for $\gamma = 4/3$ and $5/3 $ respectively.
 The above gives standard deviation  $ \sigma = \sqrt{ \gamma}/2 <p(r)> $ with
 coefficient of variance CV $ \equiv  \sigma/<p> =  \frac{\sqrt{\gamma}}{2} =
 0.57$ for an adibatic index $\gamma = 4/3$ and a  value of CV = $ 0.64 $ for
 $\gamma = 5/3 $ . Thus 
the fluctuations are about $57 \%$ to $64 \%$ of the degeneracy pressure.   
 This value is w.r.t radial ( spatial) coordinate hence it shows the 
amplitude (or rms ) of normalised fluctuations which are oscillating 
harmonically w.r.t time, in the otherwise static equilibrium configuration of
 the star. 
This results in dispersion in the pressure values in the 
background configuration and also feeds the perturbations. 

Most of the equations of state for neutron stars, estimate the  energy density
 at the core to be non-relativisitic
 ( for neutrons ), so that degeneracy pressure $p \sim n p_F^2/m  = n 2 E_F = 
2 n k_B T_F $, with $ p_F  $ being the
fermi momemtum  of neutrons, $m,n $ being the mass and number density and 
$ E_F , T_F$ being the fermi energy and fermi temperature respectively. As one
 can clearly see the fermi temperature 
fluctuations will follow on similiar lines that of the pressure from the 
relation for CV and standard deviation is around 34.2 MeV for a background
 fermi temperature 60 MeV.
 We term these fluctuations of fermi temperature as the Langevin temperature
associated with the noise, and one can see that it has a radial
dependence as well for the model here.
 It is to be noted that the statistical average of
these random fluctuations will be vanishing, as is required by a Langevin 
noise.

From equation (\ref{eq:nd}) we see that it can be written in full form
of a fluctuation dissipation relation as 
\be
 N^{\alpha \beta \zeta \delta }(\vec{x}, t,t' )  =  \frac{\gamma}{4} 
q^{\alpha \beta } q^{\zeta \delta } <p(r)>^2 e^{-i \omega(t-t')} 
\ee

 It would be further interesting to work out non-equilibrium and
dynamical configurations via the Einstein-Langevin equations and obtain
 corresponding results.  

 It is with such strong magnitudes of  fluctuations of pressure  
in the interiors of the relativisitic star, that these can
  perturb the configuration in terms of $h_{ab} $ and $\xi^a $,  by
dissipating their fermi energy. 

Other way to look at this result is due to Kubo's formalism for linear 
response theory, where any external agency causing  perturbations in the 
system, is equilivalent to stochastic
 fluctuation inside the system causing the same effect, in absence of the 
external agency \cite{kubo}. Hence stochastic  fluctuations inside the neutron 
stars of such high magnitudes that we obtain in this letter, can also be
  compared as an effect of some equivalent external agency which can perturb
 the system. The F-D theorem is a two way statement, which also states that
 perturbations  caused due to an external agency in the system 
can dissipate their energy in the isolated system  ( once the external
source is removed ) in the form of such stochastic
fluctuations. From our result we see that perturbations  induced in 
the star, can adiabatically dissipate into large magnitudes of stochastic
fluctuations of pressure inside the star. This may also be an 
interesting way to study stability issues ( for radial fluctuations )
 or if there is a possiblity of kicking the static equilibrium
configuration into a dynamically non-equilibrium state.  
Such fluctuations in the interiors have the potential
to kick a dynamically equilibrium body out of its equilibrium state, at 
critical points. 
 
Whether the  system is  perturbed  due to an external agency or noise  
originating in the system, the E-L formalism  gives  access to probe
 better for  statistical properties of the unknown matter content in the
 interiors using a first principles approach. 
 
  One can thus see in this simple formulation, we have not used
 any information of the unknown matter content of the compact object, except
 for pressures, rough estimates of fermi temperature and the spacetime metric.
 Hence we do not in advance claim over
 any specifications of the interiors which are yet unknown. Rather with this 
simple approach, we try to analyse such systems in more detail in a new way.

Usually the adiabatic index is taken to be constant for a 
given star and is approximated for the outer layers, however $\gamma $ can
 have dependence on the spatial coordinates, or more specifically a
radial dependence due to different layers inside the star given by
$\gamma(r)$. This is in fact an explorable direction on adiabatic
index and composition of relativistic stars.   

To summarize this letter, we have given a fluctuation dissipation relation,
and assigned a new dissipative effect with the adiabatic index in a very crude
 model, but quite relevant as the first step. This paves way for 
 stationary and dynamical models of relatvistic stars which are more
 interesting for observational consequences, on such lines of thought. 
It is expected that in collapsing stars, the parameters of collapse can find
 association with such fluctuations of  matter fields comprising them. The role
 of fluctuations in still unknown in this regard. However indications
 towards usefulness of such an approach comes from the theory of dynamical 
systems, critical phenomena and role  that  fluctuations play in other 
areas of Physics \cite{jarzyn} .  
 This area  still awaits even the basic developments at present. 

 It is just at the beginning stages of developing statistical 
mechanics and showing importance of fluctuations for astrophysical 
systems governed by Einstein's equations that we present the  F-D  relation
here with emphasis on the adiabatic index, the first parameter of the 
star to be veiwed in this fashion.

\section*{Acknowledgements}
The author is thankful to K.D. Kokkotas, Rajesh Nayak, Deepak Dhar and
 Dipankar Bhattacharya for useful comments and discussions over the ideas
 presented here. Also acknowledegments are due to Jasjeet Bagla for
 encouraging comments over occurrence of dissipation in models of relativistic
 stars for an earlier draft on this idea. This work is part of the project 
funded  by DST India over grant no.  WoSA/PM-100/2016. 

\end{document}